\documentclass[conference, 10pt]{IEEEtran}
\IEEEoverridecommandlockouts 
\usepackage[]{graphicx}
\usepackage{caption}
\usepackage{subfig} 
\usepackage{amsmath}
\usepackage{amssymb}
\usepackage{epsfig}
\usepackage{cite}
\usepackage{color}
\usepackage{balance}
\usepackage{amsmath}
\usepackage{amsfonts} 
\usepackage{graphicx}
\usepackage{pdfpages}
\usepackage[T1]{fontenc} 
\usepackage{array}
\usepackage{url}

\usepackage{color}
\usepackage{wrapfig}
\usepackage{lipsum}
\usepackage[hidelinks]{hyperref}
\usepackage{multirow}
\usepackage{tabularx}
\usepackage{tikz}
\usepackage{nth} 
\usepackage{algpseudocode} 
\usepackage{algorithm,algpseudocode}
\usepackage{breqn}
\usepackage{todonotes}
\usepackage{verbatim}
\usepackage{cite}

\begin{document}
\title{Day-ahead Forecasts of Air Temperature}
\author{\IEEEauthorblockN{
Hewei~Wang\IEEEauthorrefmark{1},
Muhammad~Salman~Pathan\IEEEauthorrefmark{2}\IEEEauthorrefmark{3},
Yee~Hui~Lee\IEEEauthorrefmark{4}, and
Soumyabrata Dev\IEEEauthorrefmark{2}\IEEEauthorrefmark{3}
}
\IEEEauthorblockA{\IEEEauthorrefmark{1} Beijing University of Technology, Beijing, China}
\IEEEauthorblockA{\IEEEauthorrefmark{2} ADAPT SFI Research Centre, Dublin, Ireland}
\IEEEauthorblockA{\IEEEauthorrefmark{3} School of Computer Science, University College Dublin, Ireland}
\IEEEauthorblockA{\IEEEauthorrefmark{4} School of Electrical and Electronic Engineering, Nanyang Technological University (NTU), Singapore}
\thanks{This research has received funding from the European Union's Horizon 2020 research and innovation programme under the Marie Skłodowska-Curie grant agreement No. 801522, by Science Foundation Ireland and co-funded by the European Regional Development Fund through the ADAPT Centre for Digital Content Technology grant number 13/RC/2106_P2.}
\thanks{Send correspondence to S.\ Dev, E-mail: soumyabrata.dev@ucd.ie.}
\vspace{-0.6cm}
}

\maketitle

\begin{abstract}
Air temperature is an essential factor that directly impacts the weather. 
Temperature can be counted as an important sign of climatic change, that profoundly impacts our health, development, and urban planning. Therefore, it is vital to design a framework that can accurately predict the temperature values for considerable lead times. 
In this paper, we propose a technique based on exponential smoothing method to accurately predict temperature using historical values. Our proposed method shows good performance in 
capturing the seasonal variability of temperature. 
We report a root mean square error of $4.62$ K for a lead time of $3$ days, using daily averages of air temperature data. Our case study is based on weather stations located in the city of Alpena, Michigan, United States. 
\end{abstract}

\IEEEpeerreviewmaketitle

\section{Introduction}
The weather is a climatic state of the atmosphere that includes wind speed, temperature, and humidity. 
One of the factors that influences the weather the most is air temperature. 
The air temperature is an important meteorological parameter that has a direct relationship with other meteorological parameters \textit{viz.} solar radiation, air humidity, and atmospheric pressure~\cite{manandhar2018systematic}. 
Due to the increased green house emissions in the past few decades, the air temperature has continued to increase, which is a telltale sign of abnormal climatic change~\cite{wu2021ontology}. Climate change is a phenomenon that can affect many departments including health, development, and planning~\cite{brandao2019quantifying}.  Therefore, it is of paramount importance to accurately predict the air temperature values. This in turn would assist in understanding its impact on other 
meteorological parameters. Also, an accurate prediction of temperature will lead to better heating and cooling management of buildings. 
In this paper, we propose the use of triple exponential smoothing for an accurate prediction of 
air temperature. 

The main contributions of this paper include:
\begin{itemize}
    \item We present a robust framework to forecast ground-based air temperature values using historical data;
    \item In the spirit of reproducible research, we share the source-code of our approach to the community for further benchmarking\footnote{The code related to this paper is available here: \url{https://github.com/Soumyabrata/temperature-forecasting}.}.
\end{itemize}
\section{Prediction of Air Temperature}

\subsection{Related works}
In \cite{chen2018time}, the authors used SARIMA (Seasonal Autoregressive Integrated Moving Average) techniques to predict the monthly mean air temperature in Nanjing city of china from 1951–2017. The 
forecasting accuracy of the SARIMA model was acceptable for most practical purposes. In \cite{murat2018forecasting}, the daily temperature between 1980–2010 for four different European cities was forecasted. Murat \textit{et al.} used Box-Jenkins and Holt Winters seasonal auto regressive integrated moving-average to forecast the future temperature values. 
In addition to ground-based weather station data, satellite data were also explored to estimate temperature\cite{astsatryan2021air} and analyze other atmospheric events~\cite{manandhar2017correlating}. Astsatryan \textit{et al.} in \cite{astsatryan2021air} implemented several neural network architectures to predict the hourly air temperature for up to $24$ hours in the Ararat valley of Armenia. Therefore, statistical time series techniques and neural networks offer stable frameworks in forecasting ground-based air temperature.




\subsection{Our proposed approach}
We represent the air temperature values upto time $t$ as $a_1$, $a_2$, \ldots , $a_t$. We use triple exponential smoothing technique~\cite{manandhar2019predicting} to model the seasonality of the temperature values. We model the future air temperature values $a_{t+m}$  as:

\begin{dmath}
a_{t+m} =s_t+mb_t + c_{t-L+1+(m-1) \mod L},
\end{dmath}

Here, $L$ is the season length, $s_t$ is the smooth version, $b_t$ is the linear trend estimate, and $c_t$ is seasonal corrections. In this paper, we benchmark our proposed method with persistence model and average model. The persistence model assumes that the forecasted temperature value is same as latest value, indicated by $a_{t+m} = a_t$. The average model forecasts the future temperature value as the average of the historical values, indicated by $a_{t+m} = \frac{1}{t} \sum_{t}^{} a_t$.

\section{Results \& Discussions}
In this section, we provide a detailed analysis of the prediction of temperature using triple exponential smoothing.

\begin{figure*} 
\xdef\xfigwd{\textwidth}
    \centering
       \includegraphics[height=0.3\textwidth]{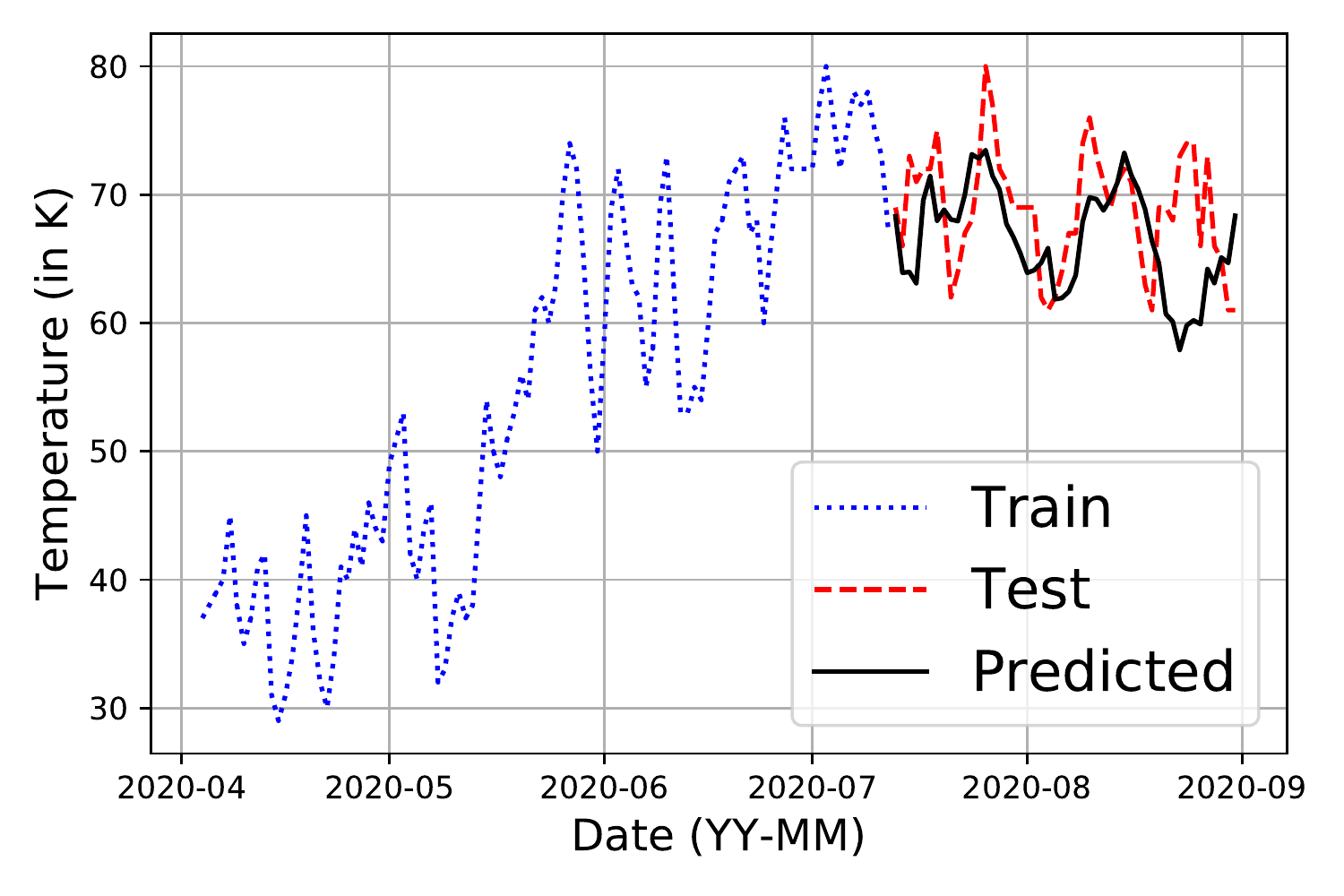}
    \label{1a}
        \includegraphics[height=0.3
        \textwidth]{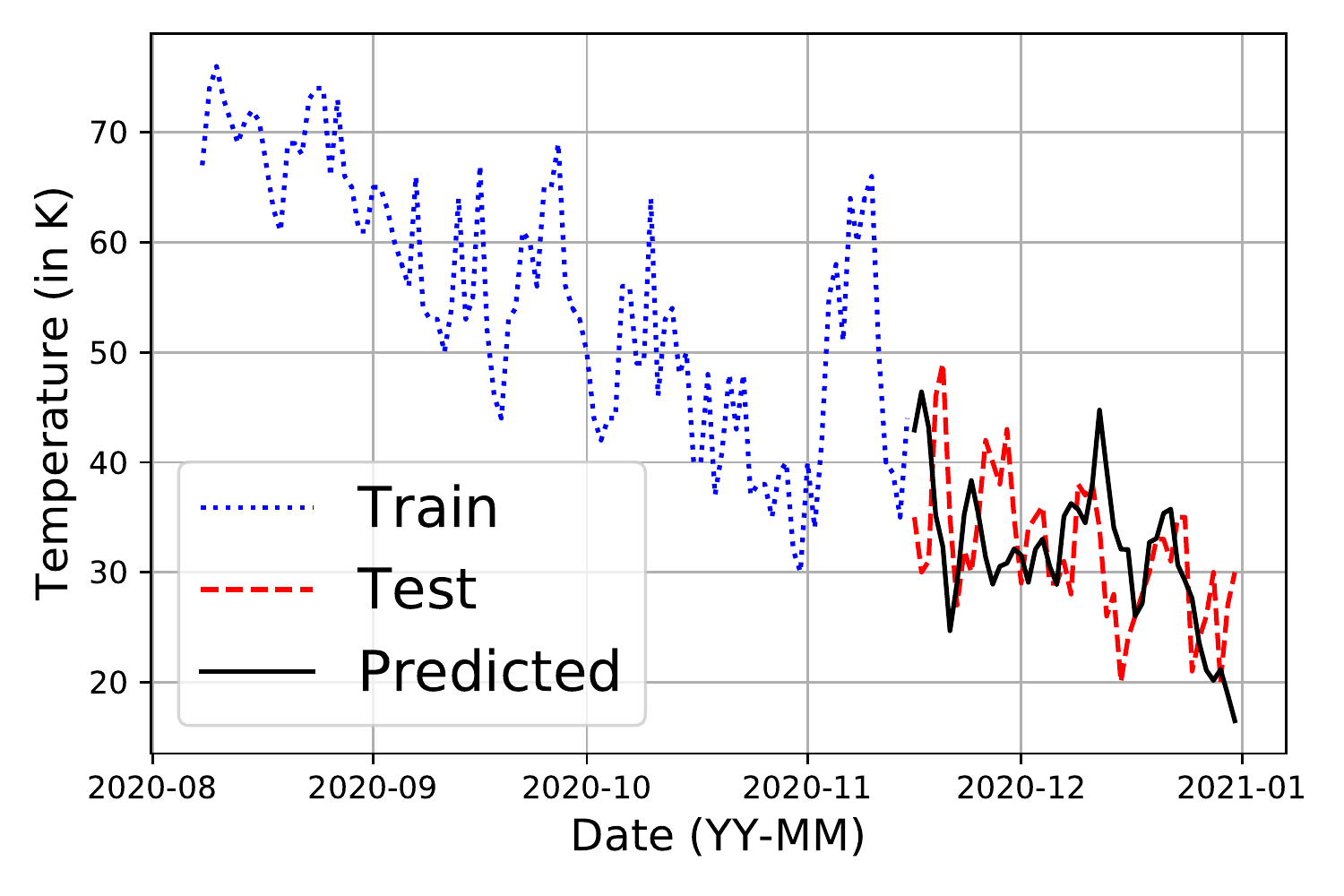}
    \label{1d} 
	\caption{We demonstrate sample illustrations of prediction of ground-based temperature. We observe that our proposed technique can clearly capture the fluctuations of the air temperature. 
	}
	\label{fig1} 
	\vspace{-0.4cm}
\end{figure*}

\subsection{Dataset}
We obtain the air temperature data from National Oceanic and Atmospheric Administration (NOAA) Climate Data Online service (CDO\footnote{https://www.ncdc.noaa.gov/cdo-web/}). We choose the weather station that is situated at Alpena Regional Airport, based in Michigan, United States. This data is the daily averaged value of the air temperature measured from the ground-based weather station. We use $6$ years worth of data for the period $2015$-$2020$.

\subsection{Qualitative evaluation}
Our proposed method can effectively capture the temperature values and provide a basis for short-term and long-term prediction. Figure~\ref{fig1} provides a subjective evaluation of our proposed method. 
We observe that our proposed method can capture the peaks and troughs of the variation of the ground-based air temperature. We use historical data of $5$ years to forecast the 
future temperature values. We observe that our proposed technique can accurately capture both the rising and falling trends of temperature values in the two subplots of Fig.~\ref{fig1}.

\subsection{Quantitative evaluation}
We compute the root mean square error (RMSE) value between the measured data and the forecasted temperature value in order to 
provide an objective evaluation of our proposed method. The performance of the temperature prediction is determined by two primary factors -- the amount of historical data for training and the length of the lead time. 
We use historical data of $5$ years in our benchmarking experiments to forecast the future temperature values. We perform $50$ distinct experiments in order to remove any sampling bias. 
We benchmark our proposed method with two popular baseline models -- persistence model and average model. 
Table~\ref{tab1} shows the RMSE value (measured in K) averaged across $50$ experiments for the benchmarking methods. 
We observed that the average model performs the worst. Our proposed method shows a consistent improvement over the persistence model for the varying lead times. The reason for this is due to the fact that temperature values remain fairly constant for shorter lead times. We also observe that the RMSE value gradually increases for larger lead times for all the methods, owing to the gradual propagation of the prediction error. 

\begin{table}[htb]
\centering
\small
\caption{We compute the RMSE values (measured in K) of the air temperature for varying lead times.}
\begin{tabular}{c|ccc}
\hline
\textbf{Lead Time} & \textbf{Proposed} & \textbf{Persistence} & \textbf{Average}\\
\hline
1 day & 3.141 & 5.320 & 15.979 \\
2 days & 4.098 & 5.669 & 14.721 \\
3 days & 4.618 & 6.819 & 15.123 \\
4 days & 5.322 & 7.835 & 16.055 \\
\hline 
\end{tabular}
\label{tab1}
\vspace{-0.3cm}
\end{table}


\section{Conclusion \& Future Work}
In this paper, we use triple exponential smoothing method for predicting future temperature values using past temperature data. Our proposed method shows better performance as compared to the other models. In the future, we intend to evaluate the impact of the length of historical data on the forecasts estimates. We also plan to further improve the forecasting performance by 
incorporating other sensor data in addition to temperature data.


\bibliographystyle{IEEEtran}

\begin{thebibliography}{1}
\providecommand{\url}[1]{#1}
\csname url@samestyle\endcsname
\providecommand{\newblock}{\relax}
\providecommand{\bibinfo}[2]{#2}
\providecommand{\BIBentrySTDinterwordspacing}{\spaceskip=0pt\relax}
\providecommand{\BIBentryALTinterwordstretchfactor}{4}
\providecommand{\BIBentryALTinterwordspacing}{\spaceskip=\fontdimen2\font plus
\BIBentryALTinterwordstretchfactor\fontdimen3\font minus
  \fontdimen4\font\relax}
\providecommand{\BIBforeignlanguage}[2]{{%
\expandafter\ifx\csname l@#1\endcsname\relax
\typeout{** WARNING: IEEEtran.bst: No hyphenation pattern has been}%
\typeout{** loaded for the language `#1'. Using the pattern for}%
\typeout{** the default language instead.}%
\else
\language=\csname l@#1\endcsname
\fi
#2}}
\providecommand{\BIBdecl}{\relax}
\BIBdecl

\bibitem{manandhar2018systematic}
S.~Manandhar, S.~Dev, Y.~H. Lee, S.~Winkler, and Y.~S. Meng, ``Systematic study
  of weather variables for rainfall detection,'' in \emph{Proc. IEEE
  International Geoscience and Remote Sensing Symposium}.\hskip 1em plus 0.5em
  minus 0.4em\relax IEEE, 2018, pp. 3027--3030.

\bibitem{wu2021ontology}
J.~Wu, F.~Orlandi, D.~O'Sullivan, and S.~Dev, ``An ontology model for climatic
  data analysis,'' in \emph{Proc. IEEE International Geoscience and Remote
  Sensing Symposium}, 2021.

\bibitem{brandao2019quantifying}
M.~Brand{\~a}o, M.~U. Kirschbaum, A.~L. Cowie, and S.~V. Hjuler, ``Quantifying
  the climate change effects of bioenergy systems: Comparison of 15 impact
  assessment methods,'' \emph{GCB Bioenergy}, vol.~11, no.~5, pp. 727--743,
  2019.

\bibitem{chen2018time}
P.~Chen, A.~Niu, D.~Liu, W.~Jiang, and B.~Ma, ``Time series forecasting of
  temperatures using {SARIMA}: An example from {Nanjing},'' in \emph{IOP
  Conference Series: Materials Science and Engineering}, vol. 394, no.~5.\hskip
  1em plus 0.5em minus 0.4em\relax IOP Publishing, 2018, p. 052024.

\bibitem{murat2018forecasting}
M.~Murat, I.~Malinowska, M.~Gos, and J.~Krzyszczak, ``Forecasting daily
  meteorological time series using {ARIMA} and regression models,''
  \emph{International agrophysics}, vol.~32, no.~2, 2018.

\bibitem{astsatryan2021air}
H.~Astsatryan, H.~Grigoryan, A.~Poghosyan, R.~Abrahamyan, S.~Asmaryan,
  V.~Muradyan, G.~Tepanosyan, Y.~Guigoz, and G.~Giuliani, ``Air temperature
  forecasting using artificial neural network for {Ararat} valley,''
  \emph{Earth Science Informatics}, pp. 1--12, 2021.

\bibitem{manandhar2017correlating}
S.~Manandhar, S.~Dev, Y.~H. Lee, and Y.~S. Meng, ``Correlating satellite cloud
  cover with sky cameras,'' in \emph{Proc. Progress in Electromagnetics
  Research Symposium-Fall (PIERS-FALL)}.\hskip 1em plus 0.5em minus 0.4em\relax
  IEEE, 2017, pp. 2166--2168.

\bibitem{manandhar2019predicting}
S.~Manandhar, S.~Dev, Y.~H. Lee, and S.~Winkler, ``Predicting {GPS}-based {PWV}
  measurements using exponential smoothing,'' in \emph{Proc. USNC-URSI Radio
  Science Meeting (Joint with AP-S Symposium)}.\hskip 1em plus 0.5em minus
  0.4em\relax IEEE, 2019, pp. 111--112.

\end{thebibliography}

\end{document}